\documentclass[12pt]{article}
\usepackage{graphics,epsfig}
\newcommand{\bce}{\begin{center}}
\newcommand{\ece}{\end{center}}
\newcommand{\beq}{\begin{equation}}
\newcommand{\eeq}{\end{equation}}
\newcommand{\bea}{\vspace{0.25cm}\begin{eqnarray}}
\newcommand{\eea}{\end{eqnarray}}

\newcommand{\ba}{\begin{array}}
\newcommand{\ea}{\end{array}}

\newcommand{\doublespace}{
    \renewcommand{\baselinestretch}{1.6}\large\normalsize}

\def\lsim{\mathrel{\rlap{\lower4pt\hbox{\hskip1pt$\sim$}}
    \raise1pt\hbox{$<$}}}     %less than or approx. symbol
\def\gsim{\mathrel{\rlap{\lower4pt\hbox{\hskip1pt$\sim$}}
    \raise1pt\hbox{$>$}}}     %greater than or approx. symbol

\def\lsim{\mathrel{\rlap{\lower4pt\hbox{\hskip1pt$\sim$}}
    \raise1pt\hbox{$<$}}}         %less than or approx. symbol
\def\gsim{\mathrel{\rlap{\lower4pt\hbox{\hskip1pt$\sim$}}
    \raise1pt\hbox{$>$}}}         %greater than or approx. symbol

\def\beq{\begin{equation}}
\def\endeq{\end{equation}}
\def\arr{\begin{eqnarray}}
\def\endarr{\end{eqnarray}}
\def\arrn{\begin{eqnarray*}}
\def\endarrn{\end{eqnarray*}}

\makeindex
\begin{document}
\vspace{2cm}

%\phantom{.}\hspace{8.0cm}{\Large \bf KFA-IKP(Th)-1995-14} \\
%\phantom{.}\hspace{11.4cm}{\large \bf 22.August 1995}
%\vspace{2cm}
\begin{center}
{\bf \huge Coulomb corrections to low energy antiproton 
annihilation cross sections on protons and nuclei.\\}
\vspace{1cm}
{\bf A. Bianconi$^{1,2)}$, G. Bonomi$^{1,2)}$
E. Lodi Rizzini$^{1,2)}$,
L. Venturelli$^{1,2)}$, A. Zenoni$^{1,2}$ } \medskip\\
{\small \sl
$^{1)}$Dipartimento di Chimica e Fisica per l'Ingegneria e 
per i Materiali, Universit\`a di Brescia, 
via Valotti 9, 25123 Brescia, Italy \\
$^{2)}$Istituto Nazionale di Fisica Nucleare,
Sezione di Pavia, Pavia, Italy \\}
\vspace{1cm}
{\bf \LARGE A b s t r a c t \bigskip\\}
\end{center}

We calculate, in a systematic way,  
the enhancement effect on $\bar{p}p$ 
and $\bar{p}A$ annihilation cross sections at low energy 
due to the initial state 
electrostatic interaction between the projectile and 
the target nucleus. This calculation is aimed at 
future comparisons 
between $\bar{n}$ and $\bar{p}$ annihilation rates 
on different targets,  
for the extraction of pure isospin channels.  
\medskip
%{\bf PACS: 25.30Fj,~24.10Eq}

%--------------------------------------------------
\newpage
\doublespace

\section{Introduction}

Recently, several sets of 
new data about antinucleon annihilation 
on nucleons and nuclei at very low energy have become 
available\cite{obe1,obe2,obe3,obe4,arm,feli,oben1}, 
and further measurements could be performed 
in the next years\cite{ad}. 
Whenever a comparison between targets or 
projectiles with different electric charge is required, 
for better understanding the underlying strong 
interaction effects (e.g. for isolating pure 
isospin contributions), it is necessary to be able to 
subtract Coulomb effects. The aim of this work 
is to calculate, as precisely and univoquely as possible, 
Coulomb effects as functions of the target mass and 
charge numbers $A,Z,$ 
and of the projectile momentum $k$ in the range 
30$-$400 MeV/c. We define $R_{A,Z}(k)$ $=$ 
$\sigma_{charged}/\sigma_{neutral}$ as the ratio between 
$\bar{p}-$nucleus 
annihilation cross sections calculated including or 
excluding Coulomb effects, at the projectile momentum 
$k$ (MeV/c) in the laboratory frame. 

In $qualitative$ sense the action 
of Coulomb effects in $\bar{p}p$ annihilations is a well 
understood process\cite{ll1,cph}. In a semiclassical 
interpretation, the electrostatic attraction acts as a 
focusing device, which deflects $\bar{p}$ trajectories 
towards the annihilation region. In quantum sense we may 
simply say that it increases the relative probability for 
$\bar{p}$ to be in the annihilation region. 
An estimation of this effect is possible by assuming 
that the actual annihilation center is pointlike and 
that there is complete independence, or factorization, 
between the effects of strong and Coulomb forces. 
Then 
$R_{A,Z}$ 
$\approx$ $\vert\Psi_Z(0)/\Psi_o(0)\vert^2$, where 
$\Psi_o(\vec r)$ is the function describing 
the free motion of a charge zero projectile, and 
$\Psi_Z(\vec r)$ describes the motion of $\bar{p}$ 
in the Coulomb field of a pointlike central charge 
$+Ze$. In this approximation 
$R_{A,Z}$ $=$ $2\pi\lambda [1-exp(-2\pi\lambda)]^{-1}$,
with $\lambda$ $=$ $Z e^2/\hbar \beta$ ($\beta$ 
is the relative velocity and $R_{A,Z}$ only depends on 
$A$ via c.m. motion within this approximation), 
and becomes $R_{A,Z}$ $\approx$ 
$2\pi\lambda$ for small velocities. Usually, at 
small velocities, the cross sections for esoenergetic 
reactions between neutral particles follow the 
$1/\beta$ law, which means constant frequency of annihilation 
events. The velocity comes in when the annihilation rate is 
divided by the incoming flux 
(perhaps suggesting that the cross section is 
not the most useful observable at very low 
energies). In the case of opposite charges for the particles 
in the initial state, the above approximation suggests 
a $Z/\beta^2$ law, at least at small $\beta$. 
However there are some limitations: 

(1) The experiments which are of interest for us 
cover a range of momenta (30$-$400 MeV/c) where 
velocities are not always small. 

(2) Proton and nuclear charges are not pointlike. 

(3) Some interplay may exist between the strong 
central potential and the action of the Coulomb forces 
that breaks the factorization of the two effects. 

(4) Some lower cutoff (in the momentum scale) must 
exist due to the action of the electron screening. 

Concerning the last point, we have attempted some calculation 
with a modified version of the codes used for the rest 
of this work. The modifications were such as 
to take into account the electron screening, with 
Thomas-Fermi distributions, for heavy nuclei. As far as 
we trust the modified codes, we don't see relevant screening 
effects at momenta $\approx$ 10 MeV/c. Apparently the 
code outputs are stable and reliable, at least at 
these kinematics and for large nuclei. 
Nevertheless the need 
to have our codes covering with precision 
very different space scales 
(atomic and nuclear, with a difference of many orders of 
magnitude) suggests a certain care. E.g., we don't get 
reliable results for larger momenta or very light nuclei 
(small variations of the parameters produce 
unstable results). So we will postpone 
a discussion on this point to the time when we have some 
alternative cross checks of these screening effects.  
Magnitude considerations anyway suggest that they 
should not be relevant at 30 MeV/c. In heavy nuclei 
the Thomas-Fermi approximation\cite{ll1} suggests 
a distance $r_B/Z^{1/3}$ between the nucleus 
and the bulk of the electronic cloud surrounding it, 
which is much larger than 1/(30 MeV/c) $\approx$ 
6 fm also for $Z$ $=$ 100. 

As we can see later (see e.g. figs. 1 and 2) the 
limitations (1) (2) and (3) are effective, and our 
results show large disagreements 
with the above $Z/\beta^2$ law, especially with 
heavy nuclei. 

In our calculations, 
the electrostatic potential has been produced by a uniform 
charge distribution with radius 1.25 $A^{1/3}$ fm. 
The annihilation is reproduced by an 
optical potential of Woods-Saxon form. 
For all but the lightest nuclei we have chosen   
zero real part, 
radius 1.3 $A^{1/3}$ fm, diffuseness 0.5 fm,  
and strength 25 MeV for the imaginary part. 
We will name this potential ``standard nuclear 
potential'' (SNP). 
The calculations have been repeated after changing 
the optical model parameters, to check for dependence 
of Coulomb corrections on these parameters (more 
details are given in the next sections). 
For the cases of 
Hydrogen, Deuteron and $^4$He targets, where low energy 
data are available\cite{obe1,obe2,obe3,obe4}, we have 
compared 
the results of the SNP with the outcome of more 
specifical (and rather different) 
potentials, which better fit the data. 

The two reasons that are behind the parameters of the 
SNP are that (i) its radius and 
diffuseness are consistent with the $A-$systematic 
parameters of the nuclear density\cite{gss}, and (ii) 
for $A$ $=$ 1 this potential 
reproduces very well the $\bar{p}p$ annihilation 
data in all of the range 30$-$400 MeV/c\cite{noi1}. 
Many other 
choices with and without a real part (both attracting 
and repulsive) or with different shapes 
can reproduce the same $\bar{p}p$ data  
(an example is given below),  
however a direct generalization of many among 
these potentials to nuclear targets is not so easy. 

Our ideal aim would be to be able to produce a curve 
$R_{A,Z}(k)$ which is independent on the specific 
potential used to simulate the strong interactions. 
For $k$ $>$ 20 MeV/c this 
is possible with very good precision in light nuclei  
and within a 10 \% uncertainty in heavy ones, as we 
show later on. Larger uncertainties are confined 
to the region of very small momenta ($k$ $<$ 20 MeV/c). 

The greatest source of interplay between the 
annihilation potential and the Coulomb interaction is the 
inversion mechanism at low energies. As widely 
discussed elsewhere\cite{noi1,pro1} 
and as seemingly measured\cite{obe2}, 
at very low energies it may happen 
that a modification of the features of the nuclear targets, 
which apparently should 
imply more effective annihilation properties, gets 
the opposite results. E.g., $\bar{p}p$ annihilation 
cross sections are larger than $\bar{p}$D and 
$\bar{p}^4$He ones at low energies. Moreover mechanisms 
(like Coulomb forces) that could be expected to  
enhance the reaction rates can loose 
effectiveness in presence of a very strong annihilation 
core. E.g., inversion is present for $k$ $<$ 200 MeV/c  
in the potential used by Br\"uckner $et$ $al$
\cite{brn1} (we name this potential BP from now onwards) 
to fit elastic $\bar{p}p$ data at $k$ $\approx$ 200 MeV/c.
The inversion property was not reported  
by these authors because, at that time, annihilation 
data at lower momenta were not available, so they didn't 
perform calculations for the inversion region. With lesser 
adjustments of the parameters, their potential 
(imaginary part: strength 8000 MeV, radius 0.41 fm 
and diffuseness 0.2 fm; corresponding parameters for the 
actractive real part: 46 MeV, 1.89 fm and 0.2 fm) 
can reasonably fit the $\bar{p}p$ annihilation data 
which have been measured in later years. However, 
it is easy to verify that any increase in the 
strength or radius of the imaginary part of their 
optical potential leads to a $decrease$ in the 
corresponding annihilation cross section 
for $k$ $<$ 200 MeV/c. In addition, putting the elastic part 
of this potential to zero leads to a twice as large 
elastic cross section. Unfortunately, since 
this potential (which has the advantage to reproduce 
elastic $\bar{p}p$ data too) 
uses radius $\approx$ 0.4 fm (for the imaginary part) 
and 1.9 fm (for the real part), and diffuseness 0.2 
fm (for both), 
its generalization to heavier nuclear targets is not  
straightforward. So with heavy nuclei we 
prefer to use the above SNP 
with standard nuclear density parameters. 
Although the inversion properties of the SNP 
are not so evident as in the BP case,  
also its outcomes are by far not $A-$linear at low momenta 
and this introduces a dependence of $R_{A,Z}$ on 
the chosen parameters. Anyway, the results that we show 
in the next section suggest that strong model 
dependence is confined to $k$ $<$ 20 MeV/c, 
even though the inversion mechanism 
is effective at larger momenta. 

Center of mass corrections have been included 
in all calculations and they are particularly 
relevant for comparison 
between $\bar{p}p$, $\bar{p}$D and $\bar{p}^4$He 
annihilations. 

Another general remark, which has a certain importance, is 
that the Coulomb focusing effect acts on the atomic scale 
and it is relevant at momenta that are smaller than the 
typical nuclear momenta. The consequence is that, if one 
uses $\bar{p}(D,p)X$ reactions to calculate the $\bar{p}n$ 
annihilation cross sections, these cross sections are 
as much Coulomb affected as the $\bar{p}p$ ones. This 
happens because the projectile is attracted by the 
deuteron charge more or less the same way, either 
it annihilates on the proton or on the neutron. So, 
while isospin invariance requires complete equality 
between $\bar{p}n$ and $\bar{n}p$ annihilation rates, 
in practics it will not be so, unless one is able to 
use free neutron targets or antiproton targets. 

\begin{figure}[htp]
\begin{center}
\mbox{
\epsfig{file=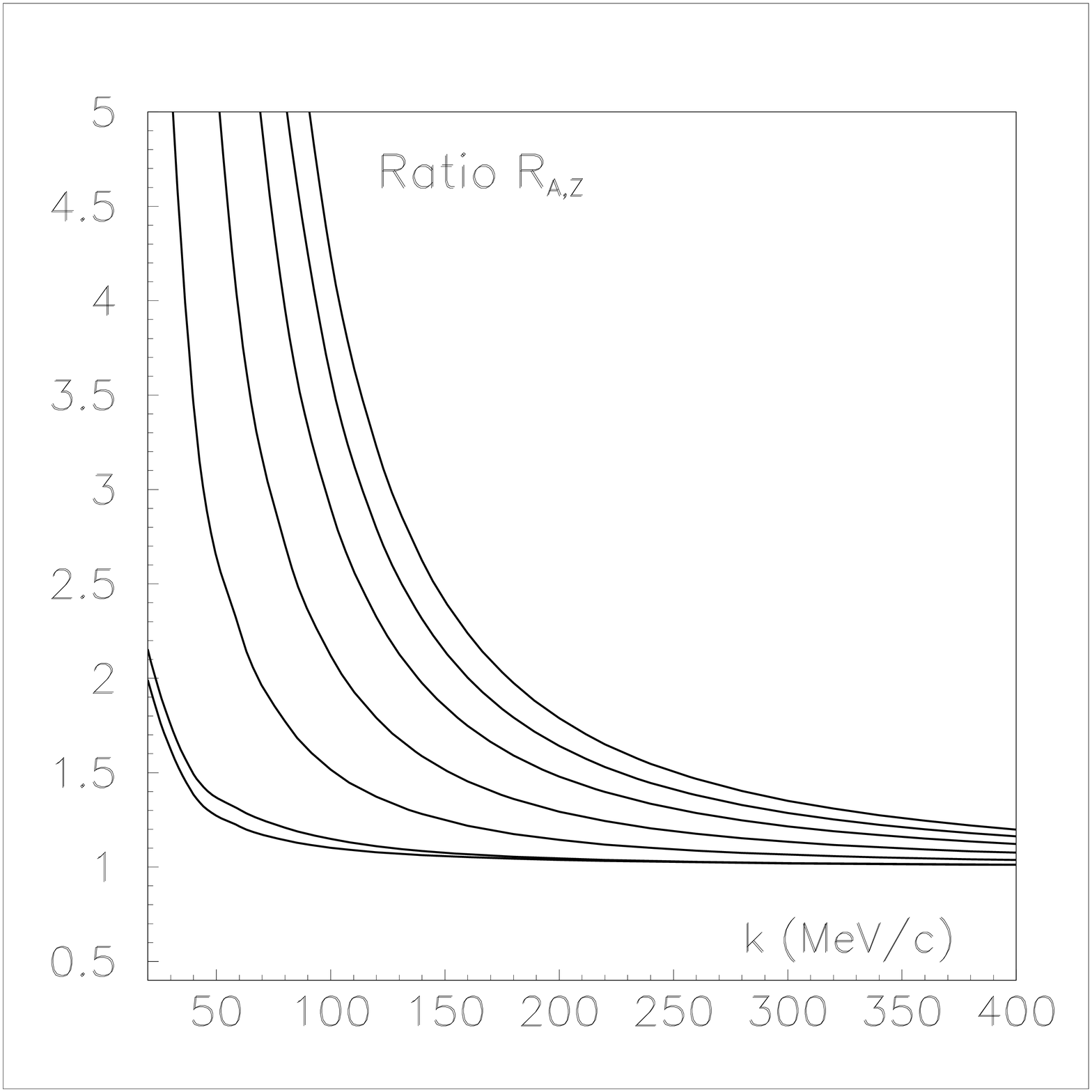,width=0.9\linewidth}}
\end{center}
\caption[]
{\small\it \label{fig1}
The ratio $R_{A,Z}$ for target nuclei: H, $^4$He, 
$^{20}$Ne and then for $A$ $=$ 50, 100, 150, 200 
and $Z$ $=$ $A/2$. Upper curves correspond to 
increasing mass number. The small difference between the 
curves relative to 
Helium and Hydrogen is due to the compensation 
between charge and center of mass effects.} 
\end{figure}

\begin{figure}[htp]
\begin{center}
\mbox{
\epsfig{file=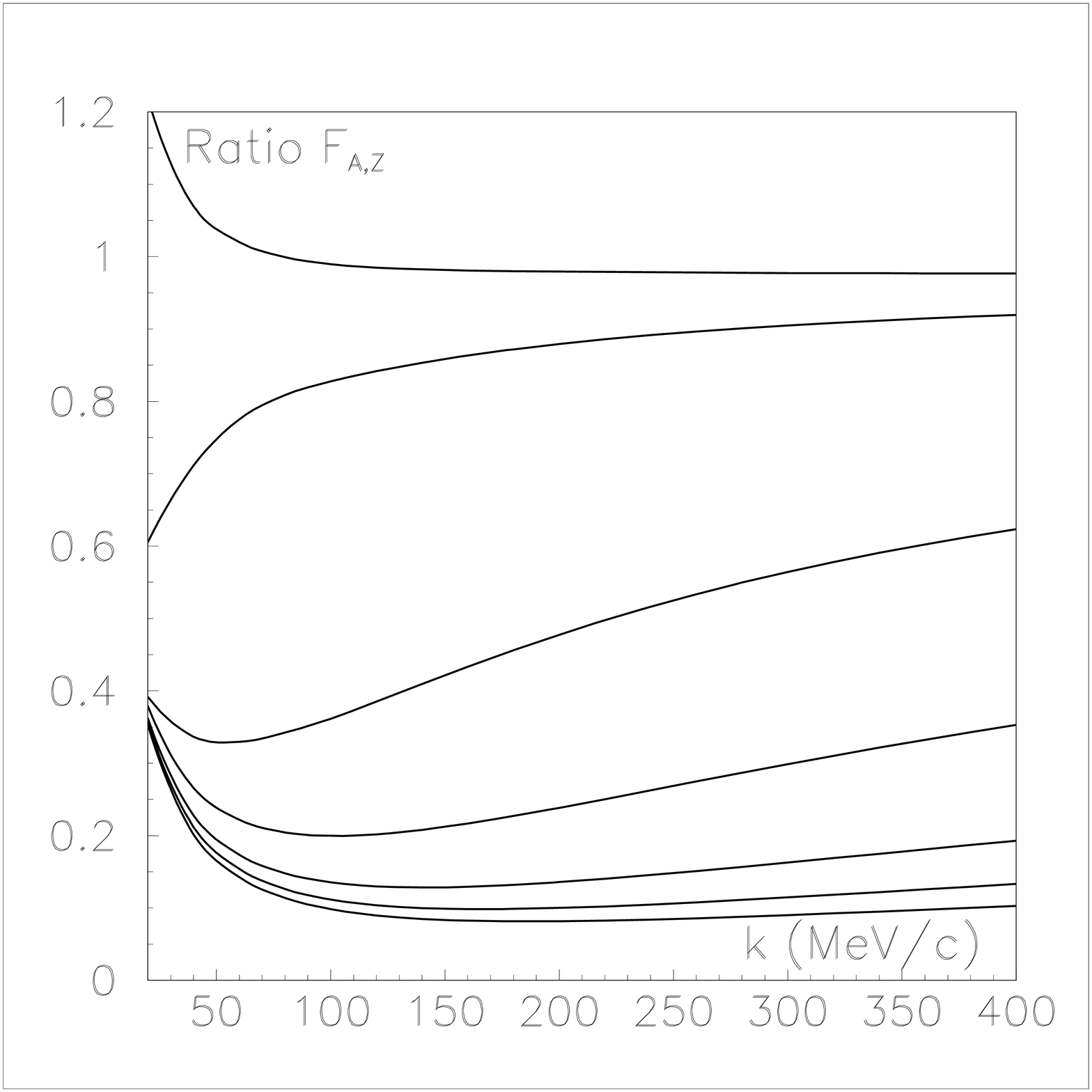,width=0.9\linewidth}}
\end{center}
\caption[]
{\small\it \label{fig2}
The ratio $F_{A,Z}$ for target nuclei: H, $^4$He, 
$^{20}$Ne and then for $A$ $=$ 50, 100, 150, 200 
and $Z$ $=$ $A/2$. Lower curves correspond to 
increasing mass number. The Hydrogen curve reaches 
the value 1 at 80 MeV/c, and is $\approx$ 0.98 over 
150 MeV/c. 
In the limit of respected pointlike prediction 
$F_{A,Z}$ should be equal to 1, so 
$Z\cdot F_{A,Z}$ 
expresses the ``effective charge'' 
of a nucleus.}
\end{figure}

\section{Qualitative trends and dependence on the 
annihilation parameters} 

In fig.1 we show the ratio $R_{A,Z}$,  
calculated with the SNP, for targets H, $^4$He,  
$^{20}$Ne, and for $A$ $=$ 50, 100, 150, 200 and 
charge $Z$ $=$ $A/2$. It can give an idea, for 
each nuclear charge, of the momentum below which it 
is not possible to neglect Coulomb effects anymore. 

Since $R_{A,Z}$ changes by orders of magnitude 
at low momenta, 
a more reasonable quantity to be 
used to verify the dependence 
of $R_{A,Z}$ on the annihilation parameters  
is the ratio $F_{A,Z}$ between $R_{A,Z}$ and its 
``pointlike'' prediction 
$R^{(p)}_{A,Z}$ $=$ $2\pi\lambda [1-exp(-2\pi\lambda)]^{-1}$,
with $\lambda$ $=$ $Z e^2/\hbar \beta$. 
This ratio, shown in fig.2 for the same target nuclei 
of fig.1, is interesting in itself, because its 
deviations from $F_{A,Z}$ $=$ 1 give an idea of the 
separation between the pointlike approximation and 
the actual nuclear behavior (notice: 
as we have tested, if one limits 
the pointlike approximation to the factor $2\pi\lambda$ 
things change little). Not accidentally, the 
``pontlike'' approximation is much worse in 
heavy nuclei. It is not so bad as far as 
the $k-$dependence is concerned, whereas it overestimates 
much the role of the nuclear charge. 
Indeed, in a wide range of momenta,  
we can write $R_{A,Z}$ $\propto$ 
$Z_{eff}(k)/k$, where the effective charge $Z_{eff}(k)$  
has a relatively slow dependence on $k$ and 
becomes, at increasing $A$, much smaller than the 
real electric charge. The fact that with a proton 
or Helium target the pointlike approximation is good 
for $k$ $>$ 100 MeV/c is of little relevance: as one 
can deduce by looking at fig.1, for light nuclei 
the charge has no role at these momenta. 

A look at a log-log plot of the annihilation 
cross sections versus $k$ with and without 
electric charge for heavy nuclei 
($A$ $=$ 50, 100, 150 and 200,  
in fig.3) shows that the ``neutral'' cross section 
is the one that behaves in the most  
unpredictable way: it has a very small k-dependence 
for 30 MeV/c $<$ $k$ $<$ 300 MeV/c, and turns  
to a $k^{-1}$ law at some $k$ $<$ 20 MeV/c. In the 
region of $k-$independence these cross sections 
are roughly proportional to $A^{2/3}$, but become less 
$A-$dependent at decreasing momenta, in agreement 
with the described inversion. 
For $k$ between 100 and 300 MeV/c the ``charged'' 
annihilation conforms 
to a rough $k^{-1}$ law, and for smaller $k$ to 
something like $k^{-1.7}$ or $k^{-1.8}$.  
``Charged'' and ``neutral'' 
cross sections  are roughly 
proportional to a similar factor 
$Z^\gamma$ or, in other words, $A^\gamma$,
with $\gamma$ close to one.  
We notice that, if 
the charge were fully effective, the most obvious 
predictions, alternative to the optical potential 
model, would suggest 
a proportionality comprised between $ZA^{1/3}$ 
and $ZA^{2/3}$ for the ``charged'' cross sections, 
and between $A^{1/3}$ and $A^{2/3}$ for the ``neutral'' 
ones; 
the first law corresponds to the S-wave geometrical 
approximation $\sigma_{ann}$ $\sim$ 
$R_{nucleus}/k$, assuming  
imaginary scattering length $\approx$ $R_{nucleus}$;  
the second law is the Distorted Wave Impulse 
Approximation, where the nuclear cross 
section is more or less the sum of the cross sections 
of those nucleons lying on the nuclear surface. 
In all models, at $k$ large enough, the charge effect 
should disappear. 

With approximation 10 \% (or slightly worse), for 
nuclei from intermediate to heavy we have found that 
it is possible to write 
$\sigma_{ann}(\bar{p}A)\cdot\beta/Z$ $\approx$ 10 mb, 
for 100 MeV/c $<$ $k$ $<$ 300 MeV/c. For 
10 MeV/c $<$ $k$ $<$ 100 MeV/c a corresponding law 
is $\sigma_{ann}(\bar{p}A)\cdot\beta^\alpha/Z^{3/2}$ 
$\approx$ 7 mb, with $\alpha$ $=$ 1.7$\div$1.8. 
Of course, in these $Z$ and $Z^{3/2}$ dependences,  
charge and mass effects mix. In the fitting formulas 
of the next paragraph the roles of $A$ and 
$Z$ will be clearly separated (for the needs of the 
application to heavy nuclei with $Z$ $<$ $A/2$). 

Only at very low momenta $\bar{p}-A$ annihilation 
cross sections 
follow the expected $k^{-2}$ law. We have compared  
annihilations on nuclei with doubled momentum, i.e.  
$k$ $=$ 2 MeV/c versus $k$ $=$ 4 MeV/c and so on. 
With the $k^{-2}$ law fully enforced, the corresponding 
ratio of annihilation cross sections should be 4. 
With Hydrogen target, 
$\sigma_{ann}(k)/\sigma_{ann}(2 k)$ $=$ 4 
within four figures for $k$-$2k$ $=$ 1-2 MeV/c, 
3.65 for 10-20 MeV/c, 3.4 for 15-30 MeV/c 
and 3.2 for 20-40 MeV/c. Things are better 
with heavy nuclei: With $A$ $=$ 200 and $Z$ $=$ 100 
we get 3.85 at 20-40 MeV/c. 

This suggests that calculations of  
scattering lengths and similar low-energy quantities, 
based on the presently available annihilation data, and  
where Coulomb effects are subtracted via the pointlike 
approximation, should be at least compared with optical 
potential analogous calculations.  

\begin{figure}[htp]
\begin{center}
\mbox{
\epsfig{file=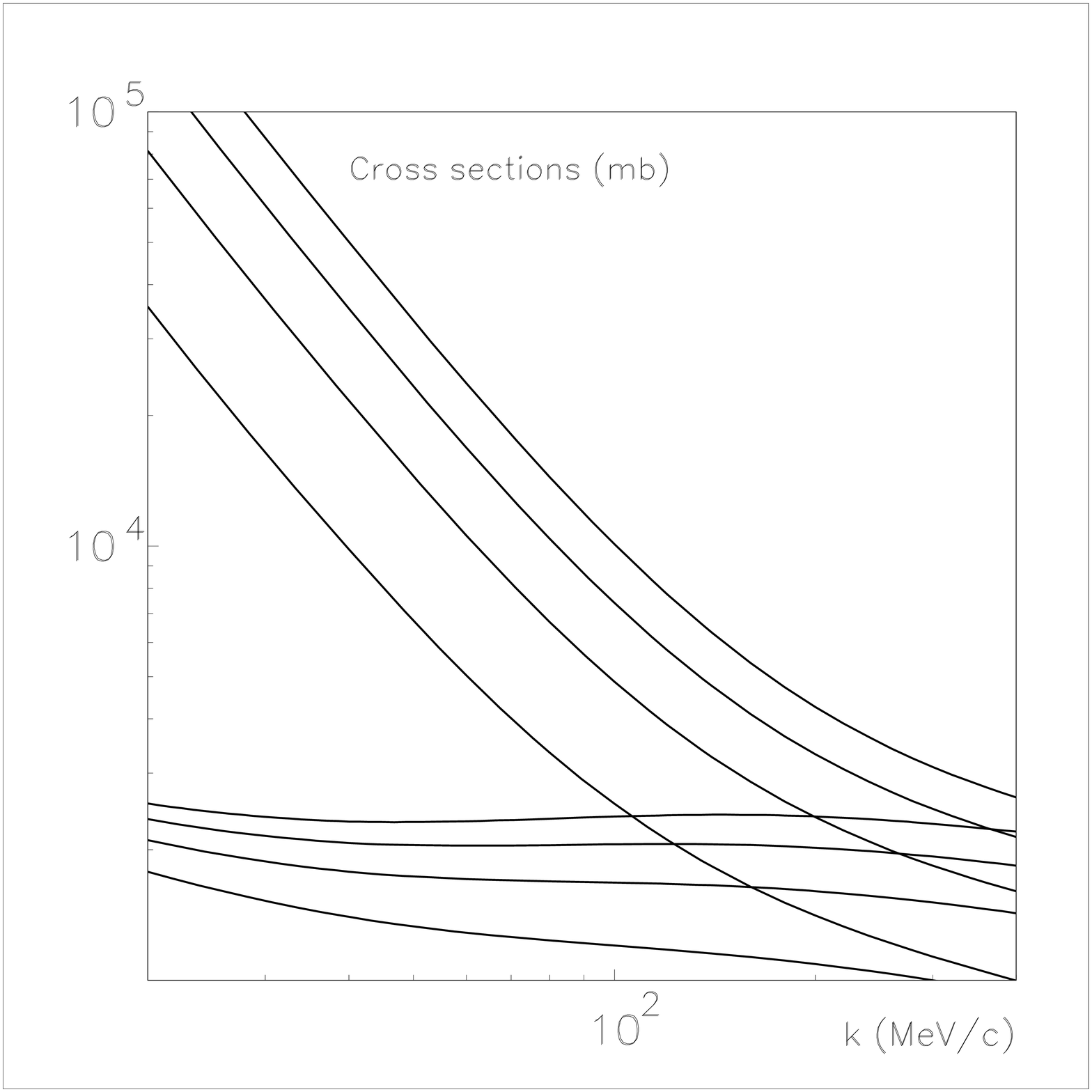,width=0.9\linewidth}}
\end{center}
\caption[]
{\small\it \label{fig3}
The four upper curves represent 
$\bar{p}A$ annihilation cross sections calculated 
within SNP including Coulomb effects, in the range 
of $\bar{p}$ momenta from 20 to 400 MeV/c. The lower 
curves reproduce the same, but without Coulomb effects. 
In both cases, larger cross sections correspond to 
increasing mass numbers. 
As well known, straight lines in log-log plots 
indicate power relations of the kind $y$ $=$ 
$x^\alpha$.}
\end{figure}

A last observation is that in fig.1 the ratio $R_{A,Z}$ 
is almost identical, despite the charge difference, 
for Hydrogen and $^4$He targets. This is due to 
the compensation between center of mass momentum 
shift and Coulomb focusing. Not considering the center of 
mass transformation can lead to large errors in 
the interpretation of light nucleus data. 

Concerning the problem of the dependence of $R_{A,Z}$ 
on the nuclear potential parameters, 
we must distinguish between the case of light 
and heavy nuclei. In the first case we have some 
low energy data 
that allow for preparing ad hoc potentials which, although  
perhaps artificially (e.g. via repulsive interactions),  
permit a reasonable reproduction of the available data. 
This allows us, with any specific light nuclear target,  
for a comparison between the outcome of 
the SNP and the outcome of a pretty different potential. 
This comparison does not show any relevant 
model dependence for $R_{A,Z}$, as showed in detail in 
the following. With heavy nuclei 
we have no alternatives to the SNP, so that comparisons  
have been performed by simply attempting some 
changes in the SNP strength and diffuseness 
and comparing the outputs. With this procedure, we 
can estimate a 
model dependence of $R_{A,Z}$ below 10 \% for heavy nuclei 
at momenta around 30 MeV/c, half of it at 50 MeV/c 
and a satisfactory model independence over 100 MeV/c. 

\begin{figure}[htp]
\begin{center}
\mbox{
\epsfig{file=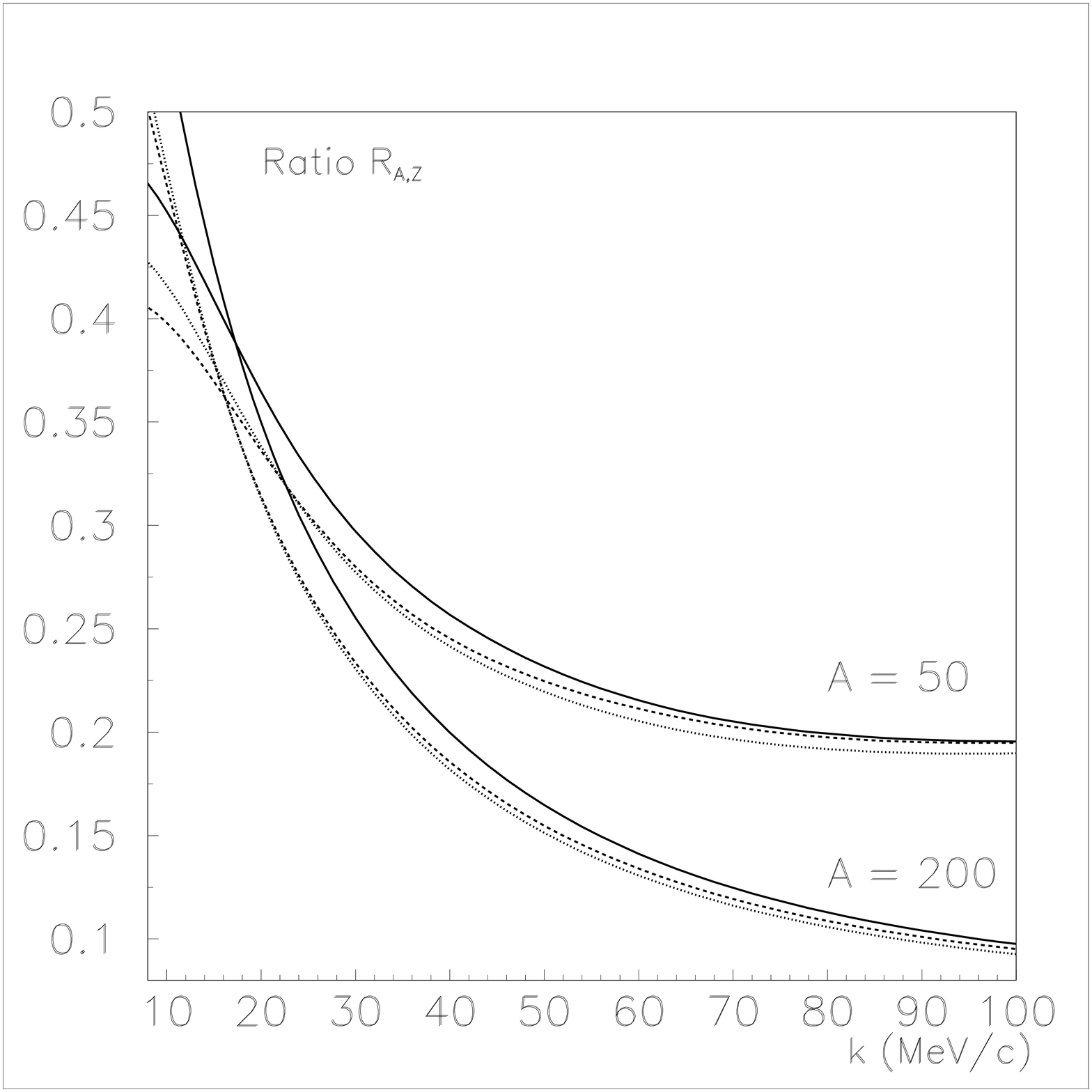,width=0.9\linewidth}}
\end{center}
\caption[]
{\small\it \label{fig4}
The ratio $R_{A,Z}$ for target nuclei 
with $A$ $=$ 50 and 200 and $Z$ $=$ $A/2$. 
For each nucleus we have used three different 
(pure imaginary and Woods-Saxon like) potentials, 
all with radius 1.3 $A^{1/3}$ fm and 

\noindent
(i) $W$ $=$ 25 MeV, $a$ $=$ 0.5 fm (continuous line),

\noindent
(ii) $W$ $=$ 12.5 MeV, $a$ $=$ 0.5 fm (dashed line), 

\noindent
(iii) $W$ $=$ 18 MeV, $a$ $=$ 0.6 fm (dotted line).
}
\end{figure}

This is clearly showed in 
fig.4, where we present $F_{A,Z}$ for the cases 
$A$ $=$ 50 and 200, $Z$ $=$ $A/2$, 
comparing the results for the three choices 
(i) $W$ $=$ 25 MeV, $a$ $=$ 0.5 fm, 
(ii) $W$ $=$ 12.5 MeV, $a$ $=$ 0.5 fm, 
(iii) $W$ $=$ 18 MeV, $a$ $=$ 0.6 fm. 
A variation of the optical potential by a factor 
two should include all the reasonable 
possibilities (much larger variations of the potential 
strength are allowed only jointly with 
compensating variations 
of the radius or of the diffuseness, which for heavy 
nuclei would not make too much sense).  
As one can clearly see in the figure, to stay 
safely within 10 \%, 
one must select momenta from 30 MeV/c upward. Below 
20 MeV/c the curves corresponding to different models 
seem to increase their separation in a 
less controllable way. 

In general, one would need 
data to impose stricter constraints on the optical 
model parameters. In fact, a qualitative 
synthesis of many attempts with different potentials, 
in light and heavy nuclei (more details on light nuclei 
are presented in the next section), suggests that  
whenever two different potentials are such as to reproduce 
similar ``charged'' cross sections, also the 
corresponding ``neutral'' cross sections will be 
similar. 
So we can say that a certain value of 
annihilation cross section is associated with a 
certain $R_{A,Z}$, whatever potential has been 
chosen to reproduce this cross section. 

\section{Fits of $R_{A,Z}$.} 

In this section we synthetize the results of the calculation 
of $R_{A,Z}$ on a wide spectrum of nuclei. We don't show
figures, since these would simply report curves 
all very similar to the previous ones. 
We give analytical fits of 
these curves, which in subintervals of 30$-$400 MeV/c 
reproduce them within specified errors. 

All the reported fits have the general form 
\arr
R_{A,Z}\ =\ 1 + C_\alpha Z \beta_{cm}^{-\alpha},
\label{fit}
\endarr 
where 
$C_\alpha$ is a constant coefficient, and 
$\beta_{cm}$ 
is the relative velocity in the center of mass frame, 
calculated via the relativistic relations between 
center of mass momentum, energy and velocity for 
a projectile with reduced mass $A M_p/(A+1)$.  
Actually, there is some $small$ difference 
between relativistic and nonrelativistic quantities 
at the larger momenta of the range only, so this 
precisation is not necessary, and one may take 
$\beta_{cm}$ $=$ $\beta_{lab}$, since at nonrelativistic 
level the relative velocity does not depend on the 
reference frame. 

With nuclei with $A$ $>$ 50 the data for $R_{A,Z}$ 
can be fit within a few percent, in the range of laboratory 
momenta 50$-$400 MeV/c, by the 
relation:  
\arr
R_{A,Z}\ =\ 1 + 10^{-5} (45 - 0.0075 A) Z \beta_{cm}^{-\alpha},
\ \ A > 50, \ 50 < k < 400 MeV/c.
\label{fit1}
\endarr
Choosing $\alpha$ $=$ 2.07 one gets a precision 
of some percent in all of the range 50$-$400 MeV/c, 
whereas choosing $\alpha$ $=$ 2.08 
the fit becomes particularly 
precise in the region 100$-$400 MeV/c 
(in practics one does not distinguish the original 
and the fitted curve anymore), precise within 10 \% 
at 70 MeV/c and within 20 \% at 50 MeV/c. 

The above fit gets worse with nuclei 
with $A$ $<$ 50. For $A$ $=$ 40 it still gives 
a 10 \% precision in the region 100$-$400 MeV/c 
(and a little worse for lower momenta). 
However a better fit (within 5 \%) is:  
\arr 
R_{40,Z}\ =\ 1 + 0.00051 Z \beta_{cm}^{-2},
\ \ 50 < k < 400 MeV/c.
\label{fitc40}
\endarr 

Following the heavy nuclei rule, the coefficient 
of $Z/\beta_{cm}^{2.07}$ would 
be 0.00041, instead of 0.00051. 
Actually the value 0.00051 is a compromise one. With 
0.00052 there is better precision 
(almost perfect superposition of curves) 
for $k$ $>$ 100 MeV/c  
and 10 \% overestimation at $k$ $=$ 50 MeV/c. 
In the region 50$-$100 MeV alone a very good fit is: 
\arr 
R_{40,Z}\ =\ 1+ 0.00084 Z \beta_{cm}^{-1.8},
\ \ 50 < k < 100 MeV/c.
\label{fitc40b}
\endarr 

For the relevant lighter nuclei, precise fits can only be 
obtained, as in the previous case,  
by systematically splitting the momentum range 
into two parts: 50$-$100 MeV/c and 100$-$400 MeV/c.  
We use the same formulas, 
with the same exponents and 
different coefficients. If we call $C_2$ and 
$C_{1.8}$ the coefficients of the $Z\beta^{-2}$ 
and $Z\beta^{-1.8}$ terms, we get: 

For $A$ $=$ 20, $C_2$ $=$ 0.00066 (which, apart from 
almost perfectly reproducing the range 100$-$400 MeV/c,  
gives a 12 \% overestimation at 50 MeV/c) and 
$C_{1.8}$ $=$ 0.97. 
It is nevertheless possible a fit of all the range 
within a few percent 
with $1+0.00088 Z\beta^{-1.83}$. 

For $A$ $=$ 12 we don't find a unique satisfactory fit 
for all of the 
required range. For the split ranges we get: 
$C_2$ $=$ 0.00080, and $C_{1.8}$ $=$ 0.0011. 
The two above coefficients allow for a precision within some 
percent. 

The coefficients for all of the nuclei with $A$ between 
12 and 40 can be interpolated quadratically exploiting the 
values given for the cases 
$A$ $=$ 12, 20, 40. 
This procedure should not introduce bigger errors than those 
ones which are related with the model dependence. 

Among the light targets, the most important case is Hydrogen,
for which we need Coulomb corrections down to 35 MeV/c. 
In this case it is possible to calculate the corrections 
related with two completely different models for the 
central potential, i.e. the SNP  
and the BP. Both produce the same annihilation 
rates in all of the considered range. 

In the first case all the range 35$-$400 MeV/can be  
fitted within 5 \% by 
\arr
R_{1,1}\ =\ 1 + C_2 \beta_{cm}^{-2},
\ \ 35 < k < 400 MeV/c,  
\label{fithy1}
\endarr
\noindent
with $C_2$ $=$ 0.00030. 
However, in the subrange 70$-$400 MeV/c the 
$\beta^{-2}$ law can be more precise, with 
$C_2$ $=$ 0.0040. This allows  for an error 
within 1 \% from 
70 to 400 MeV/c, 2 \% at 60 MeV/c and 5 \% at 50 MeV/c. 
For the subrange 30$-$70 MeV/c an almost perfect fit 
is given by the law 
\arr 
R_{1,1}\ =\ 1 + C_{1.4} \beta_{cm}^{-1.4},
\ \ 30 < k < 70 MeV/c,  
\label{fithy2}
\endarr 
with $C_{1.4}$ $=$ 0.0120. 

When one repeats the same fitting procedure starting 
with the BP for the annihilation core, 
differences are small. 
In the range 70$-$400 MeV/c the previous $\beta^{-2}$
law is as good as in the previous case, with exactly the same 
coefficient $C_2$ $=$ 0.0040. 
For the range 30$-$70 MeV/c the $\beta^{-1.4}$ law is still 
very good, with a small modification in $C_{1.4}$. 
With the previous coefficient one gets an almost 
uniform 1-2 \% overestimation of $R_{1,1}$. In this case 
the best coefficient is $C_{1.4}$ $=$ 0.0110. 
A fit within 5 \% 
of the full range 35$-$400 MeV/c is possible 
by the $\beta^{-2}$ form with $C_2$ $=$ 0.028 (instead 
of the previous 0.30). 

It must be noticed that the fact that the calculated 
value of $R_{1,1}$ is almost the same with two such 
different potentials makes this result quite reliable.  
The output of the two in terms of 
total reaction cross sections is the same, but their 
geometrical properties are completely different.  

With $^4$He and D targets the understanding of the 
structure of the annihilation potential is not very good 
yet, both because of controversial interpretation 
of data (which show strong inversion properties) and 
because it is here impossible to rely on systematical 
nuclear properties. 
As in the H case we compare fits
to the outcomes of two different potentials. 

With $^4$He we have data starting 
from 45 MeV/c. We first calculate $R_{4,2}$ 
with the SNP, that produces  
annihilation curves that don't pass too close to 
the two data points at 45 MeV/c and 70 MeV/c. 
Then we even try with a peculiar potential which, due to 
a slightly different 
annihilation core and to a repulsive elastic force, 
produces a better fit to the experimental data in the 
full range 45$-$400 MeV/c. The exact values are: 
imaginary strength 40 MeV, real (repulsive) strength 
28 MeV, radius 
1.1$\cdot 4^{1/3}$ fm, diffuseness 0.7 fm (radius and 
diffuseness are equal for the real and imaginary parts).

In the case of the SNP, 
the range 80$-$400 MeV/c can be fitted with very good 
precision by the $Z \beta^{-2}$ law with $C_2$ $=$ 
0.00130. The $Z\beta^{-1.4}$ law allows for a rather good 
fit (within 3 \%) in all of the range 40$-$400 MeV/c, 
with $C_{1.4}$ $=$ 0.0040. An improvement of the 
fit (to 1 \%) in the region 40$-$100 MeV/c can be 
obtained by the law $Z \beta^{-1.25}$ with $C_{1.25}$ 
$=$ 0.0070. 

With the second kind of potential the above $Z\beta^{1.4}$ 
fit even improves a little its accuracy (errors within 2 \% 
in all of the range 40$-$400 MeV/c). No relevant differences 
are found between the $R_{4,2}$ calculated via the two 
potentials. 

With a Deuteron target we, again, apply different choices
for the potential. First the nuclear standard
(which in the deuteron case is surely not adherent to 
the physical situation) and then a completely different 
one (which better reproduces the lowest energy 
$\bar{p}-$deuteron data) 
with imaginary strength 750 MeV, repulsive real strength 
400 MeV, real and imaginary radius 0.1 fm, real and imaginary 
diffuseness 0.6 fm. In practics the latter one 
is an exponentially 
decaying potential, having radius much smaller than 
diffuseness. 

With the SNP we can perfectly fit $R_{2,1}$ 
in the range 30$-$200 MeV/c with the $\beta^{-1.4}$ law, 
with $C_{1.4}$ $=$ 0.0060. The same fit can be extended to 
the region 200$-$400 MeV/c with error within 1 \%. In 
the latter region a  
better fit coefficient would be 0.0040 (this does not make 
a relevant difference, but with this 
coefficient the fitting law can be extended to much larger 
momenta). 
With the other potential, nothing changes. To be more precise, 
the calculated cross sections at $k$ $<$ 200 MeV/c are 
rather different at momenta below 200 MeV/c, 
but $R_{2,1}$ is the same in both cases. 

The above comparisons between couples of pretty different 
potentials confirm that for light nuclei the calculation of 
$R_{A,Z}$ is, at all practical purposes, model independent. 

\section{Summary and conclusions}

To summarize, in the full range 30$-$400 MeV/c 
we are not able to give a simple and general 
law for the Coulomb corrections, of the kind of the one 
derived from the approximation of a pointlike annihilation 
center. 
We have shown that such an approximation is 
rather poor in this momentum range.  
We have 
given analytical approximations, within reported errors, 
for the calculated values of the Coulomb correction 
with several relevant target nuclei: H, D, $^4$He, 
and then $A$ $=$ 12, 20, 40, 50, 100, 150, 200, and 
variable $Z$. By interpolation it 
should be possible to reproduce Coulomb corrections for 
most nuclear targets, starting from our formulas. 
These analytical approximations are all of the 
form $R_{A,Z}$ $=$ $1 + C_\alpha Z \beta^\alpha$, 
with $\alpha$ ranging from 1.25 to 2.08 and 
$C_\alpha$ $<<$ 1. For light nuclei (H to $^4$He) 
they should be reliably model independent, while for 
heavier nuclei it is safer to assume a residual 10 \% 
dependence on the details of the specific model used 
for describing the annihilation process.

%= = = = = = = = = = = = = = = = = = = = = = = = = = = = = = = = = = = =

\newpage

\end{document}